\documentstyle[12pt,epsf]{article}
\def\l{\label}
\def\r{\ref}
\def\be{\begin{equation}}
\def\ee{\end{equation}}
\def\bea{\begin{eqnarray}}
\def\eea{\end{eqnarray}}
\def\ni{\noindent}
\def\nn{\nonumber \\}
\def\om{\omega}
\def\eps{\epsilon}
\newcommand{\as}{\alpha_s}

\newcommand{\bef}{\begin{figure}}
\newcommand{\eef}{\end{figure}}

\newcommand{\labgraph}[1]{%
   \begin{picture}(9,9)
       \put(0,0){\makebox(9,9)%
{\centering\epsfxsize=10.0cm\leavevmode\epsffile{#1.eps}}}
   \end{picture}}

\newcommand{\fancyfig}[4]{
\begin{figure}
\begin{minipage}{14.0cm}
\begin{center}
\begin{picture}(10,10)
\put(0,8.5){#4}
\put(9,1.0){#3}
\put(1,1){\labgraph{#1}}
\end{picture}
\end{center}
\caption{#2}
\label{fig:#1}
\end{minipage}
\end{figure}}

\newcommand{\mathfigure}[3]{%
\bef
\begin{center}
\leavevmode
\hbox{%
\epsfxsize=5in
\epsfysize=7in
\epsffile{#1.eps}}
\end{center}
\caption[#3]{#2}
\label{fig:#1}
\eef}

\begin{document}
\titlepage
\begin{flushright}
DESY-96-103    \\  
MC-TH-96/18  \\
June 1996
\end{flushright}
\begin{center}
\vspace*{1.0cm}
{\Large{\bf Further analysis of the BFKL equation with momentum cutoffs}}
\end{center}
\vspace*{.75cm}
\begin{center}
M.F.McDermott$^1$   \\
and \\
J.R.Forshaw$^2$  \\
$^1$ DESY, Theory Group, Notkestr. 85, 22607 Hamburg, Germany \\
$^2$ Dept. of Physics and Astronomy, The University of Manchester, 
Manchester, England M13 9PL \\
\end{center}
\vspace*{1.2cm}
\begin{abstract}
In this paper we investigate the effect of introducing
transverse momentum cutoffs on the BFKL equation. We present solutions
in moment space for various models of the BFKL kernel for different
combinations of these cutoffs. We improve on previous calculations by using
the full BFKL kernel (rather than simplified analytic approximations).
The significance  of the next-to-leading or ``higher twist'' 
terms in the kernel are assessed.
We find that, while these terms are
negligible in the absence of cutoffs, introducing an
infra-red cutoff markedly enhances their significance.


\end{abstract}
\newpage

\section{Introduction}

The BFKL equation \cite{bfkl} can be formulated as an integral equation 
which determines the evolution (in $x$) of the unintegrated gluon distribution 
function $f(x,k^2)$. It is expected to be relevant for sufficiently small $x$,
(i.e. $\alpha_s \ln 1/x \sim 1$)
providing the, as yet unquantified, 
sub-leading corrections are not too large.
Cross-sections which are sensitive to the small-$x$ dynamics are
constructed by convoluting $f(x,k^2)$ with the appropriate coefficient
functions (impact factors). In this paper we consider the integrated
structure function
\begin{equation}
\int_0^{Q^2} \frac{dk^2}{k^2} f(x,k^2) \equiv G(x,Q^2),
\label{eq:G}
\end{equation}
which is equivalent to the usual DGLAP \cite{DGLAP} gluon density, 
$x g(x,Q^2)$, in
the double leading logarithm approximation. However, this is not a physical
observable and computation, for example, of the deep inelastic structure functions 
$F_2(x,Q^2)$ and $F_L(x,Q^2)$ requires less trivial integrals over $k^2$.

Here we do not consider the phenomenologically more interesting cross-sections.
Rather, we wish to elucidate more on the technical details which are involved
in computing $G(x,Q^2)$ using a modified BFKL kernel. Our modifications
correspond to restricting the virtualities of the $t$-channel gluons in the
real emission part of the BFKL kernel, in line with possible constraints
imposed by energy conservation and the desire to keep away from the region of
low virtualities, i.e. {\it we impose fixed infra-red and ultra-violet cut-offs
on the real emission part of the kernel}. We introduce these phase-space
restrictions on the full BFKL kernel. 

First recall the (angular averaged) BFKL equation for the evolution of the
(unintegrated) gluon density. It may may be written

\bea
f (x, k^2) & = & f^{(0)} (x, k^2) + \frac{3
\as k^2}{\pi}  \int_{x}^{x_0} \frac{ dx^{\prime} }{ x^{\prime}  }
\int_0^{\infty} \frac{ d k^{\prime2}}{ k^{\prime2} } \times \nn 
& & \left\{ {\displaystyle \frac{ ( f( x^{\prime},k^{\prime2} )
- f( x^{\prime},k^2 )  ) }{ \mid k^{\prime2} - k^2 \mid}
+ \frac{ f( x^{\prime},k^2) }{ \sqrt{ k^4 +
4k^{\prime4} } } } \right\}. 
\label{eq:bfkl}
\eea

\ni Although this equation has been known for some time there are various 
theoretical difficulties in determining its phenomenological implications.
The equation is derived in a limit where $\alpha_s$ is kept fixed, and
only includes leading terms (i.e. terms with equal powers of
$ \alpha_s $ and $ \ln (1/x) $). Furthermore, the equation involves
an integration over {\it all transverse momenta} 
including the lowest values where perturbation theory is expected to break down
(because large radiative corrections lead to  $\alpha_s > 1$ ). This
region may be regulated by introducing an infra-red cutoff \cite{akms,cl,hr}.
In addition, energy conservation demands that an ultra-violet cutoff is
present \cite{fhs}. The purpose of this paper is to determine the effect
that these cutoffs have on the mathematical structure of the equation
and its solution. 

Clearly in order to produce a realistic 
model of a small
$x$ physical cross section some attempt must be made to parameterize the region
of phase space which we have removed (below the infra-red cutoff) 
 where confinement 
effects are expected to become important. 
Some phenomenological attempts have been made in this direction \cite{akms,hr}. 
In this paper we restrict ourselves to a consideration
of what we feel may be reliably calculated within the framework of 
perturbative QCD 
and make no attempt to discuss what happens outside of this region.

The equation is best solved by taking Mellin moments with 
respect to $x,k^2$ defined by 

\begin{eqnarray}
  f(x,k^2) &=&
             \int^{c+i\infty}_{c-i\infty} \frac{d\omega}{2\pi i}
             (k^2)^{\omega+\frac{1}{2}} \tilde{f}(x,\omega) \label{eq:mt} \\
  \tilde{f}(x,\omega) & = &
               \int_{0}^{\infty} d{k^2}(k^2)^{-\frac{3}{2}-\omega}
               f(x,k^2) \label{eq:imt}                        \\
  \tilde{f}(x,\omega)  & = & \int^{c+i\infty}_{c-i\infty}
                \frac{dN}{2\pi i}x^N {\cal F}(N,\omega) \label{eq mom}   \\
  {\cal F}(N,\omega) & = & \int^{1}_{0}dx x^{-1-N}
               \tilde{f}(x,\omega).
\end{eqnarray}

The contour in eq.(\ref{eq mom}) lies parallel to the
imaginary axis and to the left of all singularities in the $N-$plane. 
In eq.(\ref{eq:mt}) the contour also lies parallel to the 
imaginary axis and to the left of all singularities associated with the
small $k^2$ behaviour of $f(x,k^2)$ but to the right of singularities
associated with its large $k^2$ behaviour. This choice ensures that the
respective inverses of these transforms exist and are consistent.
The addition of $1/2$ to the power of $k^2$
included in eq.(\ref{eq:mt}) is merely for convenience. With this
definition the $\om$-plane contour lies along $\Re e(\om) = 0$.

The following solution to the BFKL equation (in \lq Double Mellin Transform'
space) is then obtained: 

\begin{equation}
 {\cal F}(N,\omega)
= {\displaystyle \frac{{\cal F}^{(0)}(N,\omega)}{(1+N^{-1}K(\omega))} }.
\label{eq:fnnw}
\end{equation}

\ni where $K(\om)$ is the (Mellin transform of the) BFKL kernel 
\be
K(\om) = \bar{\alpha}_s [- 2\gamma - \psi(\frac{1}{2} + \omega) -
\psi(\frac{1}{2} - \omega) ] = \bar{\alpha}_s \chi(\om),
\label{eq:fulker}
\ee
\noindent $\psi$ and $\gamma$ are the logarithmic derivative of the gamma
function and the Euler constant respectively and $\bar{\alpha}_s
\equiv 3 \alpha_s / \pi$. The full kernel has an infinite set of simple
poles at $\omega$ = \{ $
\pm 1/2, \pm 3/2, \cdots $~\} as a result of the poles in the $\psi$
functions. Fig.(\ref{fig:rekern2}) shows the real part of the kernel 
in the region of interest closest to the contour.

\mathfigure{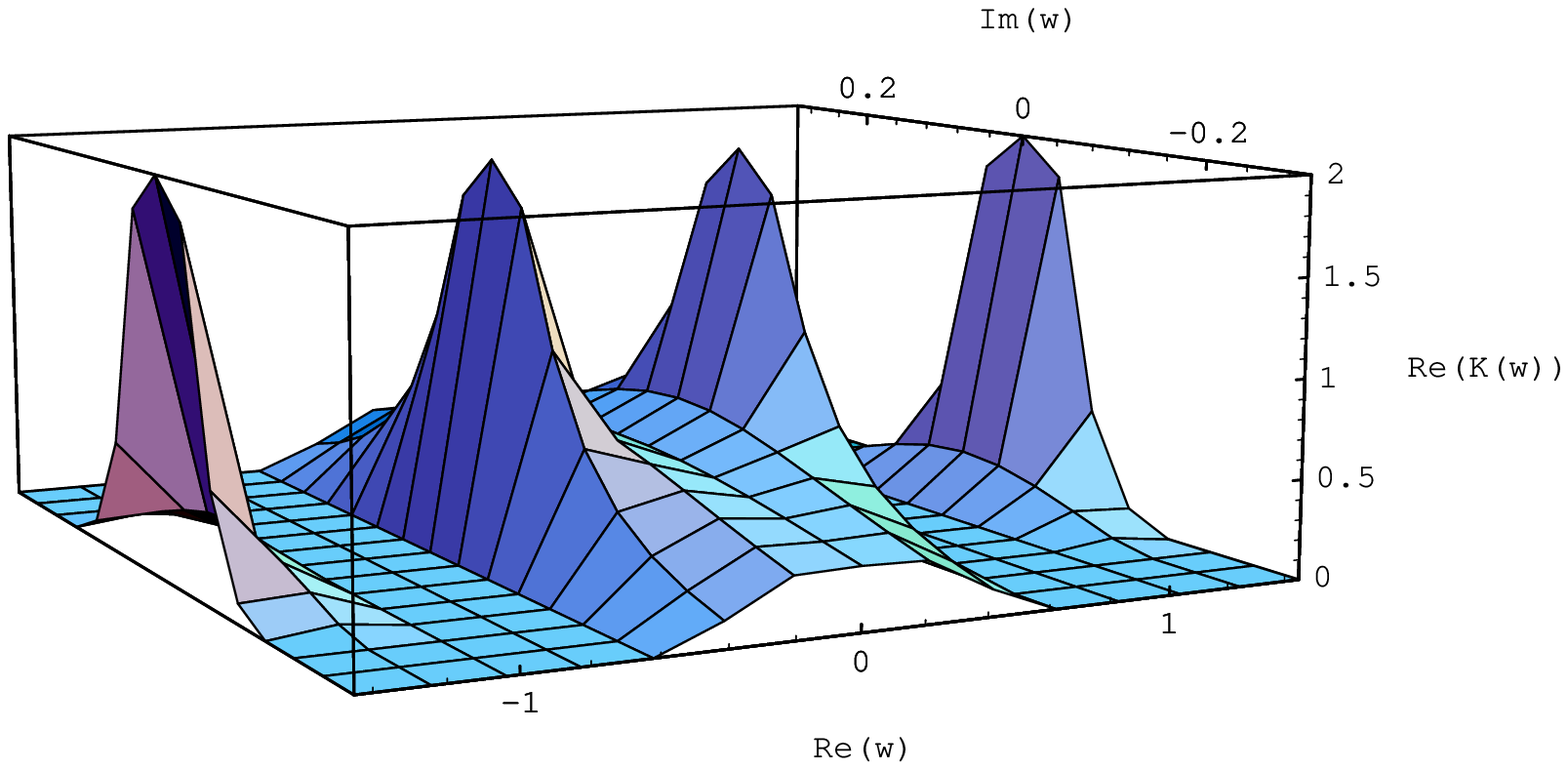}{Real part of the BFKL kernel for $\bar{\alpha}_s=0.2$.
 Note the saddle point at $ \om = (0,0) $ and the poles at
 $ \om = \pm \frac{1}{2}, \pm \frac{3}{2}, \cdots $. The poles in 
 ${\cal F}(N,\omega)$ in the $\om$-plane
 are given by the solution to $ N + K(\om) = 0 $.}
{Real part of the BFKL kernel extended to show the next-to-leading poles}



In a previous paper \cite{fmr} we presented a method for finding ${\cal F}(N,\omega)$ in the presense of infra-red, $Q_0^2$, and ultra violet, $Q_1^2$,
transverse momentum cutoffs on the real terms (i.e. the $f(x,k^{2\prime})$
terms) in eq.(\ref{eq:bfkl})) 
for a general input distribution ${\cal F}^{(0)}(N,\omega)$. The
results may be summarised as follows:

\begin{equation}
 {\cal F}(N,\omega)
= {\displaystyle \frac{1}{(1+N^{-1}K(\omega))}
({\cal F}^{(0)}(N,\omega) +
{\cal S}(N,\omega) )}.
\label{eq:gensol}
\end{equation}

The function ${\cal S}(N,\omega)$ depends in general on $Q_0^2$ and $Q_1^2$,
and upon the parameterisation chosen for the input distribution. 
Explicitly we considered three cases: the non-cutoff case, 
the single (infra-red) cutoff case and the
double (infra-red and ultra-violet) cutoff case (hereafter denoted
with a subscript $n$, $s$ and $d$, respectively).
 ${\cal S}_s (N,\omega)$
is uniquely determined by the fact that it must cancel all poles from
${\cal F}_s (N,\omega)$ to the right  of the contour in the
$\om$-plane. Similarly ${\cal S}_d (N,\omega)$ must cancel all poles
to the right and to the left of the contour. 
The structure of the solutions in $(x,k^2)$-space
depend on the pole structure in Mellin space, the positions of which
are determined by the zeros of the denominator of eq.(\ref{eq:gensol}):
\begin{equation}
N +  K(\om) = 0.
\label{eq:poles}
\end{equation}

In this paper we present solutions in $(N,k^2)$-space for various models
of the kernel and cutoff combinations. Initially we consider a
simplified model for the kernel,  
introduced by Collins and Landshoff \cite{cl}, in which only
the two poles nearest to the contour in eq.(\ref{eq:mt}) are kept.
These are the poles which lead to the dominant behaviour of $G(x,Q^2)$ in
the limit $Q^2 \to \infty$.
We present solutions for a particular input distribution, chosen such
that its 
parameters are distinct from those subsequently used for the
cutoffs on the BFKL evolution. We repeat the analysis implementing
the exact kernel of eq.(\ref{eq:fulker}), again keeping only the leading
$\omega$-plane poles. 
The integration over the $k^2$ is
performed in $N$-moment space and
the results are presented in terms of the moment of the integrated
gluon distribution $G(N,Q^2)$ \cite{cat1,frt} for each case. 
We present the effective small-$x$ slope in each scenario.

The inclusion of additional (``higher twist'') $\omega$-plane poles in the 
kernel is then implemented (i.e. the two nearest poles on each side of the 
contour in eq.(\ref{eq:mt})). Their numerical significance (as a function of 
$x$ and $Q^2$) is assessed. Again we use the full BFKL kernel.

For the case with no cutoffs present we 
find that the small-$x$ behaviour, even for fairly 
low values of $Q^2$, is dominated by the 
leading pole in the $\om$-plane (the next-to-leading poles contribute
less than $1\%$ for $x < 10^{-3}$). 
That we can safely neglect all poles other
than the two which lie nearest the $\om$-plane contour is essential for the
validity of the DGLAP formulation of the small-$x$ structure functions
(e.g. see \cite{cat1,frt,cat2,bf,km}). 

In the presence of infra-red cutoff the addition of the subleading
pole (there is only one pole to add for an infra-red cutoff) has a more 
significant effect.
We present results on the size of this effect as a function of $x$ and
$Q^2$ and comment on its possible significance. We find that the
next-to-leading terms may be as large as $\approx 20 \%$ for moderate
values of $Q^2$.

\section{The simple model revisited}

Let us briefly recap the examples given in \cite{fmr}.
The following two pole model for the kernel was used
\begin{equation}
K(\omega) = {\displaystyle \frac{  4 \ln 2 \bar{\alpha}_s}
{(1-4\omega^2)} = \frac{K_0}{(1-4\omega^2)} }.
\label{eq:clker}
\end{equation}

\ni for which eq.(\ref{eq:poles}) has only two solutions which we denote $\pm
\om_{N1}$, where
\bea
 \omega_{N1} & = & \frac{1}{2} \sqrt{1 + \frac{K_0}{N}} \label{eq:clomn} \\
\frac{1}{ 1+N^{-1} K(\om) } &  = &  
\frac{\om^2 - \frac{1}{4}}{ \om^2 - \om_{N1}^2 }.
\label{eq:tp}
\eea

In the infra-red cutoff case ${\cal S}_s (N,\omega)$ removes the pole
at $\om = \om_{N1}$, i.e. it satisfies 

\be
{\cal F}^{(0)}(N,\om_{N1}) + {\cal S}_s(N,\om_{N1})  = 0.
\l{eq:cond1}
\ee

\ni So, the solution is given by 
\begin{eqnarray}
  {\cal F }_s (N,\omega) & = & {\frac {1}{1 + N^{-1}K(\omega)}} \times
\nonumber \\
 &  & \left( {\displaystyle {\cal F}^{(0)}(N,\omega) +
 {\frac{ {\cal F}^{(0)}(N,\omega_{N1}) (\frac{1}{2} - \omega_{N1})
           Q_0^{-2(\omega -  \omega_{N1})} }
                              {  \omega-\frac{1}{2}  } } } \right)
\label{eq:fsnw}
\end{eqnarray}

\ni for $k^2 > Q_0^2$ and zero otherwise. 

In the double cutoff case ${\cal S}_d (N,\omega)$ removes both poles
in eq.(\ref{eq:tp}), i.e. it also satisfies 

\be
{\cal F}^{(0)}(N,-\om_{N1}) + {\cal S}(N,-\om_{N1})  = 0,
\l{eq:cond2}
\ee

\ni and the solution is given by

\begin{eqnarray}
{\cal F}_{d}(N,\omega) & = &
{ \frac{1}{1 + N^{-1}K(\omega)} } \times \nonumber \\
 &  & \left[ {\cal F}^{(0)}(N,\omega)  +
{\frac{1}{\Delta(\omega_{N1})}} \left\{ \Delta(\omega_{N1}){\cal S}(N,\omega)
 \right\} \right],
\label{eq:fdnw}
\end{eqnarray}
for $Q_0^2 < k^2 < Q_1^2$ and zero otherwise, where
\begin{eqnarray}
\Delta(\omega_{N1}){\cal S}(N,\omega)
& = &
{\displaystyle
\frac{ Q_0^{-2\omega} }{ ( \omega - \frac{1}{2} ) } } \nonumber \\
 & \times  &
\left[
(\frac{1}{2} + \omega_{N1})
Q_1^{2\omega_{N1}}{\cal F}^{(0)}(N,\omega_{N1}) -
(\frac{1}{2} - \omega_{N1})
Q_1^{-2\omega_{N1}} {\cal F}^{(0)}(N,-\omega_{N1}) \right] \nonumber  \\
 & + &
{\displaystyle
\frac{ Q_1^{-2\omega} }{ ( \omega + \frac{1}{2} ) } } \nonumber \\
 & \times &
\left[
(\frac{1}{2} - \omega_{N1})
Q_0^{2\omega_{N1}}{\cal F}^{(0)}(N,\omega_{N1}) -
(\frac{1}{2} +  \omega_{N1})
Q_0^{-2\omega_{N1}}{\cal F}^{(0)}(N,-\omega_{N1}) \right], \nn
& & \label{eq:deltaess} \\
\Delta(\omega_{N1}) & = & {\displaystyle
                        {\frac
                        {\frac{1}{2} + \omega_{N1}}{\frac{1}{2} - \omega_{N1}} }
                        (Q_1^2/Q_0^2)^{ \omega_{N1}}
                        - {\frac
                        {\frac{1}{2} - \omega_{N1}}{\frac{1}{2} + \omega_{N1}} }
                        (Q_1^2/Q_0^2)^{-\omega_{N1}} }
\label{eq:delta}
\eea

In \cite{cl} a simple powerlike input distribution in $k^2$ 
(truncated at the scale of the limits of the evolution integration 
over $k^2$) is used, i.e. the inputs are 

\bea
f^{(0)}_s(x,k^2)  & = &
Ax^{-\eps}(k^2)^{ ( \om_0 + 1/2 ) } \Theta(k^2-Q_0^2)  \label{eq:f0sxk} \\
{\cal F}_s^{(0)} (N,\om) & = &
\frac{A Q_0^{2(\om_0 - \om)} }{ (N + \eps) (\om_0 - \om)  }  \label{eq:f0snw}  \\
f^{(0)}_d(x,k^2)  & = &
Ax^{-\eps}(k^2)^{ ( \om_0 + 1/2 ) } \Theta(k^2-Q_0^2) (1-
\Theta(k^2 - Q_1^2)) \label{eq:f0dxk} \\
{\cal F}_d^{(0)} (N,\om) & = &
\frac{ A (Q_0^{ 2(\om_0 - \om) } - Q_1^{2(\om_0 - \om)} ) }{ (N + \eps) 
(\om_0 - \om) }.   \label{eq:f0dnw} 
\eea

The dimensionless input distribution has normalisation $A$
(which carries dimensions $[ - (\om_0 + 1/2)]$), $\om_0 < 1/2 $
ensuring that its momentum distribution dies off at large 
$k^2$ and $\eps \sim 0.08$ gives an $x$ dependence 
motivated by the observed slow rise of the proton-antiproton 
total cross section at high energies (see e.g. \cite{dl}). 
With these choices of input distribution 
eqs.(\ref{eq:fsnw},\ref{eq:fdnw}) agree with those
derived in \cite{cl} upto a minus sign (for the latter) as described
in \cite{fmr}.

For later convenience we rewrite eq.(\ref{eq:deltaess}) as

\bea
\Delta(\omega_{N1}){\cal S}(N,\omega)
& = &
{\displaystyle \frac{A}{N+ \eps} } \left( 
\frac{ Q_0^{-2\omega} X_1(\om_{N1}) }{ ( \omega - \frac{1}{2} ) } +
\frac{ Q_1^{-2\omega} X_2(\om_{N1}) }{ ( \omega + \frac{1}{2} ) }
\right) \label{eq:xdef}
\eea

\ni where are $X_1,X_2$ are given by the terms in square brackets in
eq.(\ref{eq:deltaess}) with the $A/(N+\eps)$ factor removed. 

We now investigate the sensitivity of this model to the 
degeneracy of the upper and lower scales used to parameterise the
input distribution and those employed to restrict the $k^2$ integrals 
in the evolution equation. Explicitly we replace $Q_0^2, Q_1^2$ in 
eqs.(\ref{eq:f0snw},\ref{eq:f0dnw}) with $\mu_0^2, \mu_1^2$. By keeping 
these scales distinct and using the same input in each case we
demonstrate the smooth matching between the solutions as  
$Q_0^2 \rightarrow 0, Q_1^2 \rightarrow \infty $. A comprehensive
series of analytic and numerical cross checks are 
performed which give us confidence in the solutions.

However there is a more important reason for making this distinction.
Physically we expect the input distribution to be peaked around small
values of $k^2$ with  a small tail in the perturbative high-$k^2$
region. It is these perturbative gluons which we may justifiable
evolved in perturbative QCD using the BFKL equation. 
For convenience we choose to use a theta function to model the input,
thereby  restricting ourselves to the high-$k^2$ gluons above some scale.
Hitherto we have not made the distinction between this scale
and the infra-red cutoff scale on the integral of the BFKL equation
(which defines how the input distribution evolves with $x$).
This is clearly an unphysical restriction since the choice of input
should be independent of how we choose to model the kernel.

We proceed now with the simple model in which these two scales 
are kept distinct.
For generality, we now choose to use input $ {\cal F}_s^{(0)} $ 
in the non-cutoff case too. We assume for the time being that 
$ \mu_0^2 > Q_0^2$ since we are interested in taking the limit $Q_0^2
\rightarrow 0$.

The $\om$-plane contour of eq.(\ref{eq:mt}) is closed in either the left
or right half plane according to the behaviour demanded by the
``closure factor'' in the integrand. In the non-cutoff case the
closure factor, $(k^2/\mu_0^2)^{\om}$, forces closure to the left 
for $k^2 > \mu_0^2$ and gives
\bea 
\tilde{f_{n}}(N,k^2) & =  
{\displaystyle \frac{A}{ (N + \eps) } }  & \left(   - \frac{k^{2(1/2 +
\om_0 )}}{ (1 + N^{-1} K(\om_0)) } +  \frac{k^{2(1/2 - \om_{N1})}
\mu_0^{2(\om_0 + \om_{N1})} (\frac{1}{4} - \om_{N1}^2)}{2 \om_{N1} (\om_0 +
\om_{N1}) } \right) \nn
& & \label{eq:fnnkdis} 
\eea 
\noindent and to the right for $ k^2 < \mu_0^2 $
\bea
\tilde{f_n}(N,k^2) & =  
{\displaystyle \frac{A}{ (N + \eps) } }  & \left( \frac{(\frac{1}{4} - \om_{N1}^2) k^{2(1/2 + \om_{N1})} \mu_0^{2( \om_0 - \om_{N1})}}{2 \om_{N1} (\om_0 - \om_{N1}) } \right). \label{eq:fnnkklt1dis} 
\eea

In the single-cutoff case the factor  $ (k^2/\mu_0^2)^{\om} $ enforces
closure to the left for the region $ k^2 > \mu_0^2 > Q_0^2 $ to give

\bea 
\tilde{f_s}(N,k^2) & = & 
{\displaystyle \frac{A}{ (N + \eps)} }  \left(   - \frac{k^{2(1/2 + \om_0 )}}{ (1 + N^{-1} K(\om_0)) } +  \frac{k^{2(1/2 - \om_{N1})}(\frac{1}{2} - \om_{N1})}{2 \om_{N1} }  \times \right. \nn 
 &  & \left. \left[  \frac{ (\frac{1}{2} + \om_{N1}) \mu_0^{2(\om_0 + \om_{N1})}}{(\om_0 + \om_{N1})} - \frac{ (\frac{1}{2} - \om_{N1}) \mu_0^{ 2(\om_0 - \om_{N1}) } (Q_0^2)^{2 \om_{N1}} }{(\om_0 - \om_{N1})} \right] \right) 
\label{eq:fsnkdis}
\eea
 
\noindent and to the right for $ \mu_0^2 > k^2 > Q_0^2 $
\bea
\tilde{f_s}(N,k^2) & = & 
{\displaystyle \frac{A}{(N + \eps)}\frac{ (\frac{1}{2} - \om_{N1}) \mu_0^{2( \om_0 - \om_{N1})}}{2 \om_{N1}  (\om_0 - \om_{N1}) } } \times \nn
&  &  
\left( (\frac{1}{2} + \om_{N1}) k^{2(1/2 + \om_{N1})}  - 
(\frac{1}{2} - \om_{N1}) k^{2(1/2 - \om_{N1})} (Q_0^2)^{2\om_{N1}}
\right).
\label{eq:fsnkklt1dis} 
\eea

\noindent Similarly the double cutoff case solution is given by 
\bea 
\tilde{f_d}(N,k^2) & = & \frac{A}{(N + \eps)} \left(
{\displaystyle - \frac{ k^{2(1/2 + \om_0 )} }{ (1 + N^{-1}
K(\om_0)) } } \right. \nn
& + & \left. \frac{ k^{2 (1/2 - \om_{N1} )} ( \frac{1}{2} - \om_{N1}
)  }{ 2 \om_{N1} }  \right. \nn
& \times & \left. \left[ \frac{ (\frac{1}{2} + \om_{N1}) \mu_0^{2
(\om_0 + \om_{N1})}}{(\om_0 + \om_{N1})} - \frac{ Q_0^{2\om_{N1}} X_1(\om_{N1})}{\Delta(\om_{N1})} \right] \right. \nn
& + & \left.  \frac{k^{2 (1/2 + \om_{N1} )} ( \frac{1}{2} - \om_{N1}
)  }{ 2 \om_{N1}  } \right.  \nn
& \times & \left. \left[ - \frac{ (\frac{1}{2} + \om_{N1})
\mu_1^{2(\om_0 - \om_{N1})} }
{(\om_0 - \om_{N1})} + \frac{ Q_1^{-2\om_{N1}}
  X_2(\om_{N1})}{\Delta(\om_{N1})}  \right] \right) \label{eq:fdnk}
\eea

\noindent for $  k^2  > \mu_0^2 > Q_0^2 $, 
\bea 
\tilde{f_d}(N,k^2) & = & \frac{A}{( N + \eps )}
\frac{(\frac{1}{2} - \om_{N1}) }{ 2 \om_{N1} } 
\left( \frac{ k^{2(\frac{1}{2} + \om_{N1})} (\frac{1}{2} + \om_{N1}) ( \mu_0^{ 2(\om_0 - \om_{N1}) } - \mu_1^{2(\om_0 - \om_{N1} )})} {( \om_0 - \om_{N1}
)} \right. \nn
& &
+ \left. k^{2(\frac{1}{2} + \om_{N1})} Q_1^{-2\om_{N1}} \frac{ X_2(\om_{N1}) }{ \Delta(\om_{N1}) }  - k^{2(\frac{1}{2} - \om_{N1})} Q_0^{2\om_{N1}} \frac{ X_1(\om_{N1})}{\Delta(\om_{N1})}   \right) \nn
& & \label{eq:fdnkklt1dis} 
\eea

\noindent for $ \mu_0^2 > k^2 > Q_0^2 $, and 
\bea
\tilde{f_d}(N,k^2) & = & - \frac{A}{(N + \eps)} \left(
 \frac{ k^{2 (1/2 - \om_{N1} )} ( \frac{1}{4} - \om_{N1}^2 )  }{ 2 \om_{N1}
 }  \left[ \frac{ \mu_0^{ 2(\om_0 + \om_{N1})} - \mu_1^{ 2(\om_0
+ \om_{N1})}  }{(\om_0 + \om_{N1})} \right] + \right. \nn
&  & \left. \frac{  ( \frac{1}{2} - \om_{N1} )  }{ 2 \om_{N1} }
\times \right. \nn
&   & \left. \left( k^{2 (1/2 - \om_{N1} )} Q_0^{2\om_{N1}}
\frac{X_1(\om_{N1})}{\Delta(\om_{N1})} - k^{2 (1/2 + \om_{N1} )}
Q_1^{-2\om_{N1}}\frac{X_2(\om_{N1})}{\Delta(\om_{N1})} \right) \right) \nn
& & \label{eq:fdnk2}
\eea

\noindent for $ Q_1^2  > k^2 > \mu_1^2 >  \mu_0^2 > Q_0^2  $ and
$X_1,X_2$ are defined in eq.(\ref{eq:xdef}). $\tilde{f_d}(N,k^2) $ is
the most general solution in that it uses the most general input
distribution and contains the solutions in the other cases.
 
At first the solutions derived above seem very complicated.
However one may check that they are indeed correct by 
looking at the solutions in various limits. With the definition
(\r{eq:clomn}) of $\om_{N1}$ as the positive square root 
we have that $ \Re e(\om_{N1}) > 0 $ along the contour.
We will use this result in what follows.
First we look at $f_s$ in the distinct cutoff case. If we take $
\mu_0^2  =  Q_0^2 $ in eq.(\r{eq:fsnkdis}) we obtain

\bea 
\tilde{f_s}(N,k^2) & = & 
{\displaystyle \frac{1}{ (1 + N^{-1} K(\om_0))} } 
 \left(   - \frac{A k^{2(1/2 + \om_0 )}}{ (N + \eps) }  +  \frac{A k^{2(1/2 -
\om_{N1})}(\frac{1}{2} - \om_{N1})}{(N + \eps) (\om_0 + \frac{1}{2}) } \right) 
\label{eq:fsnkdislim}
\eea

\ni which is precisely what one would obtain upon inverting  eq.(19) of \cite{cl}. 
In the limit  $ Q_0^2 \rightarrow 0 $ eqs.(\r{eq:fsnkdis},\r{eq:fsnkklt1dis})
 tend to eqs.(\r{eq:fnnkdis},\r{eq:fnnkklt1dis}). 
So we have
\bea
\lim_{\mu_0^2 \to Q_0^2}
 {\tilde{f}_s} (N,k^2,\mu_0^2,Q_0^2) & \longrightarrow & \tilde{f}_s (N,k^2,Q_0^2) \label{lim1} \\
\lim_{ Q_0^2 \to 0}
\tilde{f}_s (N,k^2,\mu_0^2,Q_0^2) & \longrightarrow & \tilde{f}_{n}(N,k^2,\mu_0^2).
\label{lim2}
\eea

\ni The double cutoff case is slightly more work but one may also prove

\bea
\lim_{\mu_0^2 \to Q_0^2 \atop \mu_1^2 \to Q_1^2 }
\tilde{f}_d (N,k^2,\mu_0^2,Q_0^2,\mu_1^2,Q_1^2) & \longrightarrow & \tilde{f}_d(N,k^2,Q_0^2,Q_1^2) \label{lim3} \\ [2ex]
\lim_{\mu_1^2 \to \infty\atop Q_1^2 \to \infty}
\tilde{f}_d (N,k^2,\mu_0^2,Q_0^2,\mu_1^2,Q_1^2) & \longrightarrow & \tilde{f}_s(N,k^2,\mu_0^2,Q_0^2) \label{lim4}.
\eea

Having seen the above cross checks working analytically we now have 
confidence in our solutions and proceed to invert the remaining 
$N$-plane integral. To complete the inversion we need to perform 
the $N$-space integral :
\be
f(x,k^2)  =  \int_{C_{N}}
                \frac{dN}{2\pi i}x^N \tilde{f}(N,k^2). \label{mtnx}  
\ee
\noindent The contour, $C_{N}$, lies parallel to the imaginary axis
and to the left of all singularities in the $N$-plane the nearest of
which is the square root branch cut at $ N = -K_0 $.  
Writing $ N= N_R + i N_I $, (with $N_R$ fixed along the contour, $C_N$) 
the $x^N$ factor may be written
\[
x^N = x^{N_R}  [ \cos ( N_I \ln (x) ) + i \sin ( N_I \ln (x) ) ]
\]
which tells us that the  integrand is oscillating along the contour 
with period $ 2 \pi / \ln x $.

Let us now consider the symmetry properties of the integrand under the transformation
$ N_I \rightarrow - N_I $. For any analytic function $f$ which is real
on some part of the real axis (which is true for the functions we will
be considering) we have by the Schwartz reflection principle: 
\bea
f(N_R + iN_I)         & = & \Re e(f(N_R + iN_I)) + i \Im m(f(N_R +i N_I)) \nn
\Re e(f(N_R -iN_I)) & = & \Re e(f(N_R + iN_I)) \nn
\Im m(f (N_R - iN_I)) & = & - \Im m(f(N_R + iN_I)). \nonumber
\eea

We see that the imaginary part of the integrand in eq.(\r{mtnx}) is odd overall
and the contributions from above and below the real axis cancel to produce a
purely real function for the gluon density. So we have

\bea
f(x,k^2) =
2 x^{N_R} \int^{\infty}_{0} \frac{d N_I}{2\pi} 
\left[ \cos (N_I \ln x )  \Re e( \tilde{f} ) - \sin ( N_I \ln x )  \Im m(\tilde{f}) \right]
\label{intosc} 
\eea

The non-trivial $N$-plane singularity structure associated with 
the  square root branch points at $ N = - K_0 $ 
and $ N = 0 $ contained in the factor $ 1 / 2\om_{N1} $ necessitates a
numerical inversion of the remaining transform.
We achieve this using a Fortran
program which calls the NAG integration routine D01ASF which is
specifically designed for integrals with oscillating integrands.
Having performed the $N$-plane integral we have explicitly checked
the smooth matching conditions for all ranges of $k^2$ and $x$.

The small-$x$ behaviour, in the non-cutoff and infra-red cases, is dominated by
the square root branch point at $N = -K_0$ (c.f. eq.(\ref{eq:clomn}) ).
In the double cutoff case the singularity structure in the $N$-plane
is governed by $\Delta(\om_{N1}) = 0$ defined in eq.(\ref{eq:delta}), 
the zeros of this function occur at pure imaginary values of
$\om_{N1}$ and also at $\om_{N1}=0$ \cite{cl,fmr}. 
In \cite{cl} it was shown that for the double-cutoff case this branch 
point singularity cancels between terms to leave a softer behaviour 
given by a sum of poles to the right of $K_0$ in the $N$-plane. 

It appears from eq.(\ref{eq:fdnk}) that we now have a 
singularity in $\tilde{f_d}(N,k^2)$ at $N = -K_0$ associated with the
$ 1/2\om_{N1} $ factor. A careful consideration of the terms in square
brackets, in the limit $\om_{N1} \rightarrow 0$,  
however reveals that this singularity is cancelled and one is
left with the sum of poles in the $N$-plane coming from the zeros of 
$\Delta(\om_{N1})$ for which $\om_{N1}$ are imaginary as before \cite{cl,fmr}.


\section{Using the full BFKL kernel} 

In this section we explain how to implement a more realistic model
which uses the analytic form for the full BFKL kernel given by
eq.(\ref{eq:fulker}) ($K(\om) = \bar{\alpha}_s \chi(\om)$).
Initially  we include only the nearest poles at $\pm 1/2$ in the
kernel. These are referred to as the leading
twist poles since, as we will see, they lead to the largest power in $Q^2$.
The pole positions in the $\om$-plane,  at $ \pm \om_{N1} $, are 
given by the solution to the following simultaneous equations 

\bea
N_R + \Re e(K(\om_{N1})) & = &  0 \nn
N_I + \Im m(K(\om_{N1})) & = &  0
\label{traneqs}
\eea

\noindent such that
\be
\left| \Re e(\om_{N1}) \right| < \frac{1}{2}
\ee

\noindent The exact position of these poles depends on the choice of position
of the contour in the $N$-plane,  and on the position along this
contour.
Looking at the graph of $ \Re e(K(\om)) $  in fig.(\ref{fig:rekern2}) 
we see that the solutions of the real part of
the equation are given by a curves of constant height, $-N_R $, such
that $ - N_R > K_0 $, where $K_0$ is the height of the saddle point at
$\om = (0,0)$ (imagine taking a horizontal slice through the saddle). 
Provided this condition is satisfied, $ N_R $ lies to the left of all singularities in
the $ N $-plane. As
$ N_I $ varies the solutions $\pm \om_{N1}$ move along these curves.
Having chosen the position of the $N$-plane contour we find $\om_{N1}$
numerically, the complex solution then feeds into the residue of the
poles calculated below.

We may now invert the solutions (\ref{eq:fnnw},\ref{eq:fsnw},\ref{eq:fdnw})
to $(N,k^2)$-space, by closing the contour in the appropriate
direction and summing over the residues of the enclosed poles as
before.
To do this we need to determine the contribution that the crucial factor, $ 1 +
N^{-1}K(\om)$, makes to the residue. For example, if a
particular term contains a closure factor forcing closure
to the left we need to know the residue of the pole at $ \om = -\om_{N1} $, so
we expand the factor about this pole to arrive at
\bea
\frac{ 1 }{ 1 + N^{-1} K(\om)} & = & \frac{ {\cal R} (\om) }{(\om + \om_{N1})} \nn
{ \cal R}^{-1}                   & = & \left. \frac{\partial}{\partial
\om}
 \left( 1 + \frac{\bar{\alpha}_s}{N}  \chi(\om) \right) \right|_{\om = - \om_{N1}} \nn
 { \cal R} (-\om_{N1})  & = &  \left. \frac{1}{ (\bar{\alpha}_s / N)
\chi^{\prime} (-\om_{N1}) } \right. \nn 
\chi^{\prime} (\om)  & = & \frac{\partial \chi (\om)}{\partial \om} 
\label{eq:resid}
\eea

\noindent with similar relations for the pole at $ \om = \om_{N1} $. 
The function $\chi (\om)$ which retains the full $\om$-dependence of 
the kernel is given by eq.(\ref{eq:fulker}) 

For the non-cutoff case, closing to the left for  
$ k^2 > \mu_0^2 $, we have
\bea 
\tilde{f_n}(N,k^2) & =  
{\displaystyle \frac{ A N }{ (N + \eps) } }  & \left(   -
\frac{k^{2(1/2 + \om_0 )}}{ (N + K(\om_0)) } +  \frac{k^{2(1/2 -
\om_{N1})} \mu_0^{2(\om_0 + \om_{N1})} }{ (\om_0 +
\om_{N1}) \bar{\alpha}_s \chi^{\prime}(-\om_{N1}) } \right) 
\label{eq:fnnfful1} 
\eea 
\noindent and to the right for $ k^2 < \mu_0^2 $,
\bea
\tilde{f_n}(N,k^2) & =  
{\displaystyle \frac{A N}{ (N + \eps) } }  & \left( \frac{ k^{2(1/2 +
\om_{N1})} \mu_0^{2( \om_0 - \om_{N1})}}{ (\om_0 - \om_{N1}) \bar{\alpha}_s 
\chi^{\prime}(-\om_{N1}) }  \right) 
\label{eq:fnnfful2} .
\eea

\noindent With an infra-red cutoff on the evolution, $ Q_0^2 $, we get 
\bea 
\tilde{f_s}(N,k^2) & =  
{\displaystyle \frac{A N }{ (N + \eps)} } & \left(   - \frac{k^{2(1/2
+ \om_0 )}}{ (N  +  K(\om_0)) } +  \frac{ k^{2(1/2 - \om_{N1})} }
{\bar{\alpha}_s \chi^{\prime}(-\om_{N1}) }  \times \right. \nn 
 &  & \left. \left[  \frac{ \mu_0^{2(\om_0 + \om_{N1})}}{(\om_0 + \om_{N1})}
- \frac{ (\frac{1}{2} - \om_{N1}) \mu_0^{2(\om_0 - \om_{N1}) } Q_0^{2(2
\om_{N1}) }}{(\frac{1}{2} + \om_{N1}) (\om_0 - \om_{N1}) } \right] \right) 
\label{eq:fsnkful1}
\eea

\noindent for $  k^2 > \mu_0^2 > Q_0^2 $. For  $ \mu_0^2 > k^2 >
Q_0^2 $ the input term is closed to the right to give
\bea
\tilde{f_s}(N,k^2) & = & 
{\displaystyle \frac{A N }{ (N + \eps) } \frac{\mu_0^{2 (\om_0 -
\om_{N1}) }}{\bar{\alpha}_s \chi^{\prime}(-\om_{N1})  } }  \times  \nn
&  &  \left( \frac{k^{2(1/2 + \om_{N1})}}{(\om_0 - \om_{N1})} -
\frac{ (\frac{1}{2} - \om_{N1})  k^{2(1/2 - \om_{N1})} Q_0^{2(2\om_{N1})}  }{(
\frac{1}{2} + \om_{N1}) ( \om_0 - \om_{N1} ) }  \right). 
\label{eq:fsnkful2}
\eea
\noindent Similarly the double cutoff case solution is given by 
\bea 
\tilde{f_d}(N,k^2) & = & \frac{AN}{(N+\eps)} \left(
{\displaystyle - \frac{ k^{ 2(1/2 + \om_0 )} }{ (N + \eps) (N +
K(\om_0)) } } +  \frac{ 1 }{ \bar{\alpha}_s \chi^{\prime}(-\om_{N1}) }
\right. \nn
& \times & \left. \left[ \frac{ \mu_0^{2(\om_0 + \om_{N1})} k^{2 (1/2 - \om_{N1}
) } }{(\om_0 + \om_{N1})} - \frac{ \mu_1^{2(\om_0 - \om_{N1})} k^{ 2
(1/2 + \om_{N1} ) } } {(\om_0 - \om_{N1})} \right] \right.  \nn
& + & \left. \frac{ k^{2 (1/2 + \om_{N1} ) } Q_1^{2(-\om_{N1})}
X_2(\om_{N1}) 
 - k^{2 (1/2 - \om_{N1} )} Q_0^{2(\om_{N1})} X_1(\om_{N1})  }
{   \bar{\alpha}_s \chi^{\prime}(- \om_{N1}) \Delta(\om_{N1})
(\frac{1}{2} + \om_{N1})} \right) \nn
& &  \label{eq:fdnkful1} 
\eea
\noindent for $  Q_1^2 > \mu_1^2 > k^2  > \mu_0^2 > Q_0^2 $, 
\bea 
\tilde{f_d}(N,k^2) & = & 
\frac{ A N }{( N + \eps ) \bar{\alpha}_s \chi^{\prime}(-\om_{N1}) }
\left( \frac{ k^{2(1/2 + \om_{N1}) }  ( \mu_0^{ 2(\om_0 - \om_{N1}) } - \mu_1^{2(\om_0 - \om_{N1} )})} {( \om_0 - \om_{N1}
)} \right. \nn
& + & \left. \frac{ k^{2(1/2 + \om_{N1})} Q_1^{2(-\om_{N1})}
X_2(\om_{N1}) - (k^2)^{1/2 - \om_{N1}} Q_0^{2(\om_{N1})}
X_1(\om_{N1}) }{ \Delta(\om_{N1}) ( \om_{N1} + \frac{1}{2} ) } \right) \nn
& & \label{eq:fdnkful2} 
\eea
\noindent for $ Q_1^2 > \mu_1^2 > \mu_0^2 > k^2 > Q_0^2 $, and 
\bea
\tilde{f_d}(N,k^2) & = &
\frac{ A N }{( N + \eps ) \bar{\alpha}_s \chi^{\prime}(-\om_{N1}) }
\left( \frac{ k^{2(1/2 - \om_{N1} )}  ( \mu_0^{ 2(\om_0 + \om_{N1})
} - \mu_1^{2(\om_0 + \om_{N1} )})} {( \om_0 + \om_{N1} )} \right. \nn
& + & \left. \frac{ k^{2(1/2 + \om_{N1})} Q_1^{2(-\om_{N1})}
X_2(\om_{N1})  - k^{2(1/2 - \om_{N1})} Q_0^{2(\om_{N1})}
X_1(\om_{N1}) 
 }{ \Delta(\om_{N1}) ( \om_{N1} + \frac{1}{2} ) } \right) \nn
& & \label{eq:fdnkful3} 
\eea

\noindent for $ Q_1^2 > k^2 > \mu_1^2 >  \mu_0^2 > Q_0^2 $. 

Note that in these equations  only terms corresponding to the input 
distribution change in the different $k^2$-regimes. So the first and
second terms in eqs.(\ref{eq:fsnkful1}, \ref{eq:fdnkful1}) are the same as
eq.(\ref{eq:fnnfful1}); the first terms in eqs.(\ref{eq:fsnkful2},
\ref{eq:fdnkful2})
are the same as eq.(\ref{eq:fnnfful2}). The third term in
eq.(\ref{eq:fdnkful1}) and the second in eq.(\ref{eq:fdnkful2}) (i.e.
those involving $\mu_1^2$) are present
because a different input is used in the double cutoff case.
The remaining terms in eqs.(\ref{eq:fsnkful1}, \ref{eq:fsnkful2}, \ref{eq:fdnkful1}, \ref{eq:fdnkful2})
represent the effect of the cutoffs on the evolution.

The remaining $N$-plane integral is now performed numerically 
as described above employing extra
subroutines in the Fortran programs to calculate $ K(\om_0) $ and $
\chi^{\prime} (\om_{N1})$. Again we verify that the numerical
solutions have the expected properties, i.e. that they
satisfy the various limits detailed above. This
has been done for $ k^2 > \mu_0^2 $ and  $k^2 < \mu_0^2 $ 
corresponding to the inversion of  eqs.(\ref{eq:fnnfful1},
\ref{eq:fsnkful1}, \ref{eq:fdnkful1}) and eqs.(\ref{eq:fnnfful2},
\ref{eq:fsnkful2}, \ref{eq:fdnkful2}) respectively and we find that all the
appropriate cross checks are observed. 

Fig.(\ref{fig:s10w12}) shows the effective small-$x$ slope,
$\lambda_{{\rm eff}}$, defined by 

\begin{equation}
\lambda_{{\rm eff}} \equiv \frac{ \partial \ln G(x,Q^2)}{ \partial \ln 1/x}.
\end{equation}

\noindent in each case, as a percentage of its asymptotic value of $ \lambda_0
= 4 \ln 2 {\bar \alpha}_s $.
Two curves are shown for the double cutoff case corresponding to the
range of variables expected from HERA kinematics. 
The two values chosen for $Q_1^2$,
 $10^3 $ and $10^5$ GeV$^2$, correspond (somewhat arbitrarily)  
to the maximum value of $Q^2$ at 
$x = 10^{-2}$ and the total centre-of-mass energy respectively.

\fancyfig{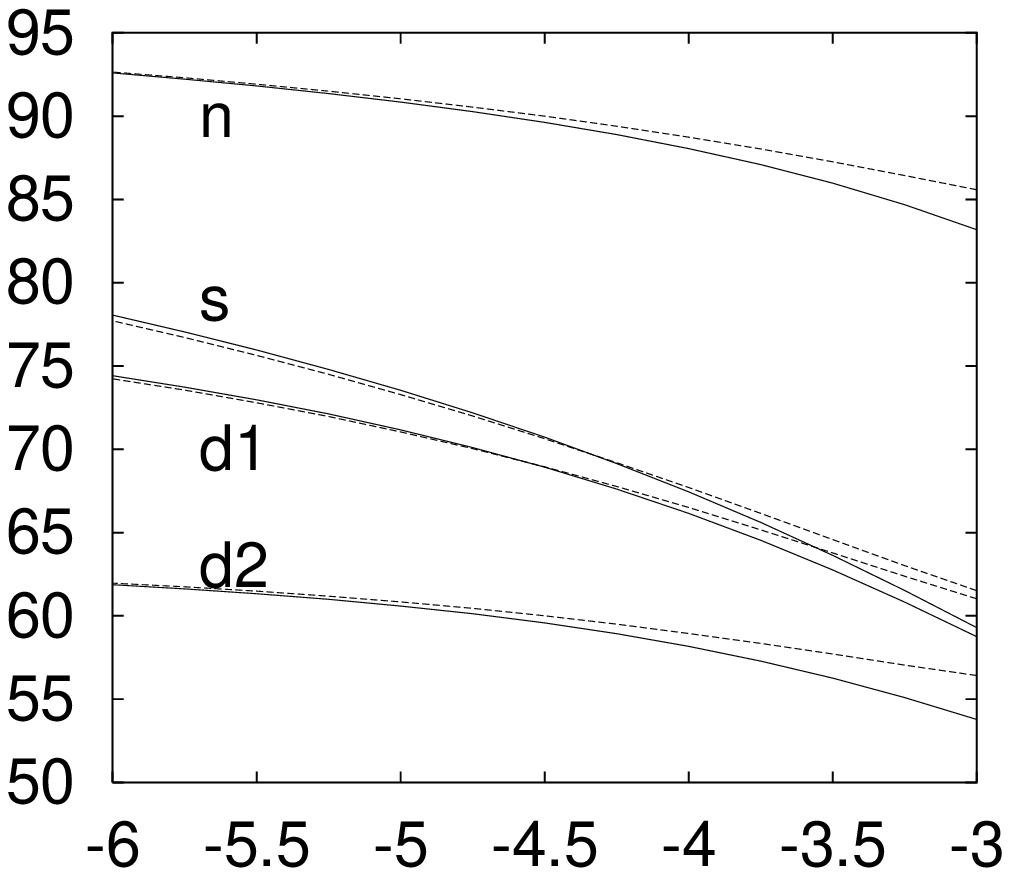}{The effective slope of $G(x,Q^2)$ as a percentage
of its asymptotic value $ \lambda_0 = 4 \ln 2 {\bar \alpha}_s $, 
for $Q^2 = 10 $ GeV$^2$. The solid line in each pair of curves
is for $\om_0 = -1.0$ and the dashed line for $\om_0 = -2.0$. 
At small enough $x$ the slopes approach a value independent of 
the specific choice of $\om_0$. For the curves shown in the 
single cutoff case (labelled '$s$')$, Q_0^2 = 0.5~$GeV$^2$.  
In the double cutoff case we show two sets of curves;
for $d1$, $Q_1^2 =10^5$ GeV$^2$ whereas for $d2$,
$Q_1^2 =10^3$ GeV$^2$. $\mu_1^2 $ is set
very high ($10^9$ GeV$^2$) so that the comparison can be made for 
the same input distribution (here $\mu_0^2 = 0.5~$GeV$^2$).}
{$\log_{10} (x)$}{$ {\displaystyle \frac{100 \lambda}{\lambda_0} }$}


\subsection{Higher twist terms in the leading $\ln (1/x)$ approximation.}

We now consider including further poles associated with the
BFKL kernel. As explained above these poles are subleading in $Q^2$
and may be referred to as ``higher twist'' terms.
Since the solutions get very complicated with more poles included, we 
restrict ourselves to the non-cutoff and  infra-red cutoff cases
as these contain the essential features.  


Keeping the nearest and next-to-nearest poles in the kernel ($w= \pm
1/2,3/2$) leads to four solutions of eq.(\ref{eq:poles}) at 
$\om = \pm \om_{N1}, \pm \om_{N2} $ where

\be
1 < \left| \Re e(\om_{N2}) \right| < \frac{3}{2},
\ee

\ni this can easily seen from fig.(\ref{fig:rekern2}) which shows the full kernel
out to $ \om = \pm 3/2 $.  We note that $\Re e(\om_{N2}) \approx
\Re e(\om_{N1}) + 1$. 

The essential features can be illustrated
 using the following simple
four pole kernel which has the same pole structure as the full
kernel close to the contour:

\be
K(\omega) =  \frac{K_1}{(1-4\omega^2)}  + \frac{K_2}{(9-4\omega^2)} .
\label{eq:ker4}
\ee

\ni where $K_1, K_2$ are determined, for example, by matching this
kernel onto the full kernel (eq.(\ref{eq:fulker})) at $\om = 0,\pm 1$.
Eq.(\ref{eq:poles}) now has four solutions given by the solutions to a
quartic equation in $\om_{N}$. 
The residue factor at these poles is given by (c.f. eq.(\ref{eq:resid}))

\bea
 { \cal R} (\om_{N}) & = & {\displaystyle \frac {(4 \om_N^2 - 1) (4 \om_N^2 - 9) }  
{ 8 \om_N \left[ K_1 (4 \om_N^2 - 9) + K_2 (4 \om_N^2 - 1)  \right] } }
\eea

\ni which becomes singular either when $\om_N = 0$ or when the term in square
brackets in the denominator becomes zero. For the leading poles at $\om
= \pm \om_{N1} $ the former condition is satisfied when 
$ N = - K_0 = - 4 \ln 2 \bar{\alpha}_s $ i.e at the original branch point. 
For the subleading poles in the $\om$ -plane at $\om_N = \pm \om_{N2}$
the latter condition may be satisfied, this occurs for two complex 
conjugate values of $ N \approx  -K_0 (0.1
\pm 0.25)$. Since these singularities lie far to the right of the branch
point in the $N$-plane at $ N = - K_0 $ one expects them to be
insignificant for small enough $x$.


This analysis may be extended trivially to include further poles in the kernel.
The  pole structure of the kernel  determines that the solutions to 
$ N + K(\om_{N1}) = 0 $ are spaced approximately one unit apart in terms 
of $ \Re e(\om_{N1}) $. This leads, on closing the $\om$ plane contour, 
to a sum of $ k^2 $ terms with each subsequent term being
approximately one power in $k^2$ down on the previous one 
(or equivalently in $Q^2$ once the $k^2$ integral has
been performed). In this sense all the terms after the first one are
higher twist in the language of the operator product expansion and may
be neglected for large enough values of $ Q^2 $. In what follows we
investigate, and attempt to quantify, the magnitude of the contribution 
made by these additional poles at intermediate values of $ Q^2 $ and
$x$, by explicitly including four poles in the BFKL 
kernel. This will be referred to as the next-to-leading pole
calculation and denoted with a superscript ``nl'' (with a
superscript ``l'' now for the leading pole solutions).

The non-cutoff case solution is straightforward. 
The effect of including this pole is to introduce terms in
eqs.(\ref{eq:fnnfful1},\ref{eq:fnnfful2}) in $\om_{N2}$ 
identical in form to the existing terms in
$\om_{N1}$, which are, approximately speaking,
of order $k^2/\mu_0^2$ relative to the leading, $k^{2(1/2-\om_{N1})}$, terms.

Following the procedure outlined in \cite{fmr}, in the infra-red
cutoff case the function 
${\cal S}_s(N,\omega)$ is determined by the fact that is removes the
right half plane poles (at $\om =  \om_{N1}, \om_{N2} $) from $ {\cal F}_s $.
Assuming a form which has the appropriate closure factor
\[
{\cal S}(N, \omega) = Q_0^{-2\om} 
{\displaystyle \left(\frac{a_1}{\omega - \frac{1}{2}} +
\frac{a_2}{\omega - \frac{3}{2} } \right) },
\label{sdc}
\]
\ni one solves the following simultaneous equations: 

\bea
{\cal F}^{(0)}(N, \om_{N1})
 + {\cal S}(N,  \om_{N1})
& = & 0  \nn
{\cal F}^{(0)}(N, \om_{N2})
 + {\cal S}(N,  \om_{N2})
& = & 0  \nonumber
\eea
\ni to give
\bea
a_1 & = & \frac{ (1/2 - \om_{N1}) (1/2 - \om_{N2})}{(\om_{N1}-\om_{N2})} \nn
& \times & \left[ (3/2 - \om_{N1})Q_0^{2\om_{N1}} {\cal F}^{(0)}(N,\om_{N1}) - 
(3/2 - \om_{N2})Q_0^{2\om_{N2}} {\cal F}^{(0)}(N,\om_{N2}) \right] \nn
a_2 & = & \frac{ (3/2 - \om_{N1}) (3/2 - \om_{N2})}{(\om_{N2}-\om_{N1})} \nn
 & \times & \left[ (1/2 - \om_{N1})Q_0^{2\om_{N1}} {\cal F}^{(0)}(N,\om_{N1}) - 
(1/2 - \om_{N2})Q_0^{2\om_{N2}} {\cal F}^{(0)}(N,\om_{N2}) \right].  \nn 
\eea

\ni We can now see that both $a_1$ and $a_2$ contain information about 
both the leading and next-to-leading poles.
As we will see, this has a
significant effect on the $x$ dependence of the solution.
 
After inverting the $\om$-plane transform as appropriate for the
values of $k^2$ (using the full kernel, $K(\om) = \bar{\alpha}_s
\chi(\om)$, of eq.(\ref{eq:fulker})) we arrive at the following solutions: 
\bea 
\tilde{f_s}(N,k^2) & = & 
{\displaystyle \frac{A N }{ (N + \eps)} }  \left(   - \frac{k^{2(1/2
+ \om_0 )}}{ (N  +  K(\om_0)) } + \frac{ k^{2(1/2 - \om_{N1})} }
{\bar{\alpha}_s \chi^{\prime}(-\om_{N1}) } \times   \right. \nn 
 &  & \left.  \left[  \frac{ \mu_0^{2(\om_0 + \om_{N1})}}{(\om_0 + \om_{N1})}
- Q_0^{2\om_{N1} } \left(
\frac{a_1}{(1/2 + \om_{N1}) } +
\frac{a_2}{(3/2 + \om_{N1}) } \right) \right]  \right) \nn 
 & + &  ( \om_{N1} \rightarrow \om_{N2}) 
\label{}
\eea

\noindent for $  k^2 > \mu_0^2 > Q_0^2 $. 
For  $ \mu_0^2 > k^2 > Q_0^2 $  we have 
\bea
\tilde{f_s}(N,k^2) & = &  
\frac{A N}{(N + \eps)} \times \left(
\frac{k^{2(1/2 + \om_{N1})} }{\bar{\alpha}_s \chi^{\prime}(-\om_{N1}) } 
\frac{ \mu_0^{2(\om_0 - \om_{N1})}}{(\om_0 - \om_{N1})}  \right. \nn
& - & \left.
\frac{k^{2(1/2 - \om_{N1})} }{\bar{\alpha}_s \chi^{\prime}(-\om_{N1}) } 
Q_0^{2\om_{N1} } \left[
\frac{a_1}{(1/2 + \om_{N1}) } +
\frac{a_2}{(3/2 + \om_{N1}) } \right] \right)  \nn 
& + &  ( \om_{N1} \rightarrow \om_{N2}) 
\eea

Until now we have been working with the unintegrated gluon
distribution function $f(x,k^2)$. This is not a physical observable. 
It must be convoluted with an appropriate coefficient function to
arrive at an observable for the  particular process under
consideration. For simplicity, we present our results in terms of the
gluon structure function $G(x,Q^2)$ defined in eq.(\ref{eq:G}). 
Since we have analytic expressions for ${\tilde f} (N, k^2)$ in each
case we may perform this $k^2$ integral prior to inverting the
$N$-plane integral.

In the non-cutoff case we split the integral up
into two regions and integrate eqs.(\ref{eq:fnnfful1},\ref{eq:fnnfful2}),
\be
{\tilde G}_n (N, Q^2) = \int^{Q^2}_{\mu_0^2} \frac{d k^2}{k^2} {\tilde f}_n
(N, k^2) + \int^{\mu_0^2}_0 \frac{d k^2}{k^2} {\tilde f}_n (N, k^2)
\ee
\noindent which gives for the  leading pole solution
\bea
{\tilde G}^l_n (N, Q^2) & = &
{\displaystyle \frac{ A N }{ (N + \eps) } }  \left(   -
\frac{ (Q^{2(1/2 + \om_0 )} - \mu_0^{2(1/2 + \om_0)} ) }{(\om_0 + 1/2) (N + K(\om_0)) } \right.\nn
& + & \left. \frac{ \mu_0^{2(1/2 + \om_0)} }{ \bar{\alpha}_s \chi^{\prime}(-\om_{N1})}
\left( \frac{1}{(1/2 + \om_{N1})(\om_0 - \om_{N1})} -
\frac{1}{(1/2 - \om_{N1})(\om_0 + \om_{N1})} \right) \right.\nn
& + &  \left. \frac{ Q^{2(1/2 - \om_{N1})} }{ \bar{\alpha}_s 
\chi^{\prime}(-\om_{N1})} \left(\frac{\mu_0^{2(\om_0 + \om_{N1})}}{(1/2 -
\om_{N1})(\om_0 + \om_{N1})} \right) \right).
\label{eq:Gn}
\eea

\ni The next-to-leading solution has an additional term relative to this

\bea
{\tilde G}^{nl}_n (N, Q^2) & = & {\tilde G}^{l}_n (N, Q^2) + 
\frac{ A N }{ (N + \eps) } \times \nn
&   & \left( \frac{ \mu_0^{2(1/2 + \om_0)} }{ \bar{\alpha}_s \chi^{\prime}(-\om_{N2})}
\left( \frac{1}{(1/2 + \om_{N2})(\om_0 - \om_{N2})} -
\frac{1}{(1/2 - \om_{N2})(\om_0 + \om_{N2})} \right) \right. \nn
&   & + \left. \frac{ Q^{2(1/2 - \om_{N2})} }{ \bar{\alpha}_s 
\chi^{\prime}(-\om_{N2})} \left(\frac{\mu_0^{2(\om_0 + \om_{N2})}}{(1/2 -
\om_{N2})(\om_0 + \om_{N2})} \right) \right).
\label{eq:GN2}
\eea

In the single-cutoff case, which has the same input, the integral is
non-zero in two regions:
\be
{\tilde G}_s (N, Q^2) = \int^{Q^2}_{\mu_0^2} \frac{d k^2}{k^2} {\tilde f}_n
(N, k^2) + \int^{\mu_0^2}_{Q_0^2} \frac{d k^2}{k^2} {\tilde f}_n (N, k^2)
\ee
\noindent the result differs from the non-cutoff solution
only in the coefficient of $Q^{2(1/2 - \om_{N1})}$
\bea
{\tilde G}^l_s (N, Q^2)
& = & {\tilde G}^l_n (N, Q^2) -  {\displaystyle \frac{AN}{(N + \eps)} }
\left( \frac{Q^{2(1/2 - \om_{N1})} }{\bar{\alpha}_s 
\chi^{\prime}(-\om_{N1})}  \frac{\mu_0^{2(\om_0 - \om_{N1})} (Q_0^2)^{2
\om_{N1}} }{(1/2 + \om_{N1})(\om_0 - \om_{N1})}  \right). \nn
& & \label{eq:Gs}
\eea

The next-to-leading solution can no longer be expressed in terms of
the leading solution plus an extra piece from the new pole as in eq.(\ref{eq:GN2})
as a result of the information about the position of {\it both} poles tied up in the
coefficients $a_1,a_2$. The solution is given by

\bea
{\tilde G}^{nl}_s (N, Q^2) & = & {\tilde G}^{nl}_n (N, Q^2) - 
    \frac{         (Q^{2(1/2-\om_{N1})} - Q_0^{2(1/2-\om_{N1})}) }
         {(1/2-\om_{N1}) \bar{\alpha}_s \chi^{\prime}(-\om_{N1}) }
\times \nn
& & Q_0^{2\om_{N1}} 
\left[ \frac{a_1}{1/2+\om_{N1}} + \frac{a_2}{3/2+\om_{N1}} \right] \nn
& - &  ( \om_{N1} \rightarrow \om_{N2}).
\label{eq:GS2}
\eea

For completeness we also give the double cutoff solution for the
leading pole case for the same $k^2$-regimes as in the single cutoff case: 

\bea
{\tilde G}_d^l (N, Q^2) & = & {\tilde G}_n^l (N, Q^2) + {\displaystyle
\frac{AN}{(N + \eps)} } \times  \left( \right. \nn
&   & \left. \frac{ Q^{2(1/2 + \om_{N1})} }{ \bar{\alpha}_s 
\chi^{\prime}(-\om_{N1})} \left( \frac{\mu_1^{2(\om_0 - \om_{N1})}}{(1/2 +
\om_{N1})(\om_0 - \om_{N1})} \right) \right. \nn
& - & \left. \frac{ Q_0^{2(1/2 + \om_{N1})} ( \mu_0^{2(\om_0 - \om_{N1})}
- \mu_1^{2(\om_0 - \om_{N1}) } ) } 
{ \bar{\alpha}_s \chi^{\prime}(-\om_{N1}) (1/2 +\om_{N1})(\om_0 - \om_{N1})} \right. \nn
& + & \left. \frac{ 1 } { \bar{\alpha}_s \chi^{\prime}(-\om_{N1}) (1/2 +
\om_{N1}) \Delta(\om_{N1})  } \right. \nn 
& \times &  \left. \left[ - \frac{ (Q^{2(1/2 - \om_{N1})} - Q_0^{2(1/2 - \om_{N1})})
Q_0^{2\om_{N1}} X_1 (\om_{N1}) }{ (1/2 -\om_{N1})}  \right. \right. \nn
&   & \left. \left. +   \frac{ ( Q^{2(1/2 + \om_{N1})} - Q_0^{2(1/2 + \om_{N1})}
)  Q_1^{-2\om_{N1}} X_2 (\om_{N1}) }{ (1/2 + \om_{N1}) } \right]
\right) .
\label{eq:Gd}
\eea

We now compare the model with two poles in the  kernel to that with
four  in order to determine the significance of these next-to-leading
poles as a function of $Q^2$ and $x$. To this end we work with the
following quantity  (which is the percentage 
difference between the two solutions)

\be
R(x,Q^2) = 100 \times \frac{ G^{(nl)} (x,Q^2) - G^{(l)} (x,Q^2) }{ G^{(l)} (x,Q^2) }.
\ee

From the above analysis we expect $R_n$ to go to zero as $Q^2
\rightarrow \infty$ since the coefficients of the leading $Q^2$ piece
(the term in $Q^{2(1/2-\om_{N1})}$) are the same in each case (see
eqs.(\ref{eq:Gn},\ref{eq:GN2}) ).
In contrast, the coefficients of the ``leading twist''
pieces for the single cutoff case are different (see
eqs.(\ref{eq:Gs},\ref{eq:GS2})) so we expect $R_s$ to
tend to a constant (which will be determined by the choice of input
parameters $\om_0,\mu_0^2,Q_0^2$; the overall normalization cancels in
the ratio). Figs.(\ref{fig:rnqw},\ref{fig:rsqw})
illustrate these  features for a specific choice of input parameters. 

Let us now consider the $x$ dependence of these solutions. We have
seen from the simple four pole model of the kernel that the
singularities in the $N$-plane associated with $\om_{N2}$ are
subleading compared to the branch point at $N = -K_0$.
The non-cutoff solution contains terms which involve either $\om_{N1}$
or $\om_{N2}$ but not both, hence one expects that the significance of
the next-to-leading poles will decrease as $x \rightarrow 0$. One can
see from fig.(\ref{fig:Rnxw}) that this is indeed the case. In the single
cutoff case however the effects of the poles are coupled in the
coefficients $a_1,a_2$, this makes the $x$ dependence of $R_s$
considerably less trivial since the coefficient of the leading piece 
in $x$  in $G_s^{(nl)} $ and $ G_s^{(l)} $ are different.

Fig.(\ref{fig:Rsxw}) shows $R_s$ as a function of $x$, for various
$Q^2$ and  two different values of the input parameter $\om_0$. $\om_0 
= -1.0$ is special in that it lies in the same interval in $\om$ as
$\om_{N2}$. The curves shown for $\om_0 = -2.0$ are
typical of those for which $\om_0$ lies outside of this interval ($<
-1.5 $). For the latter, $R_s$ is
seen to increase as $x$ decreases eventually tending to a constant
percentage (of the order of 10 \%)  as $x \rightarrow 0$. 

It seems that for intermediate values of $x$ when an infra-red cutoff 
is present the next-to-leading poles in the BFKL equation are
significant and including only one of the 
subleading poles leads to a behaviour in $x$ similar to what one 
would expect for the sum of two powers.

This conclusion follows given the fact that we have chosen to
regulate the infra-red physics using a theta function cutoff on the
integral in eq.(\ref{eq:bfkl}) and a specific form (\ref{eq:f0sxk})
for the input distribution. It would certainly be interesting to
establish whether it is true in general.

\section*{Conclusions}

We have presented analytic solutions to the BFKL equation with infra-red and
ultra-violet cutoffs in moment space, including the leading pole of 
the full BFKL kernel. We have investigated the numerical significance of the
subleading ('higher-twist') poles. We find that when one includes an
infra-red cutoff the assumption that these poles are irrelevant is
brought into question.

\section*{Acknowledgements}

One of us (M.M.) would like to thank Graham Ross for supervision and
discussions, Jochen Bartels for discussions and 
P.P.A.R.C. for financial assistance.


\fancyfig{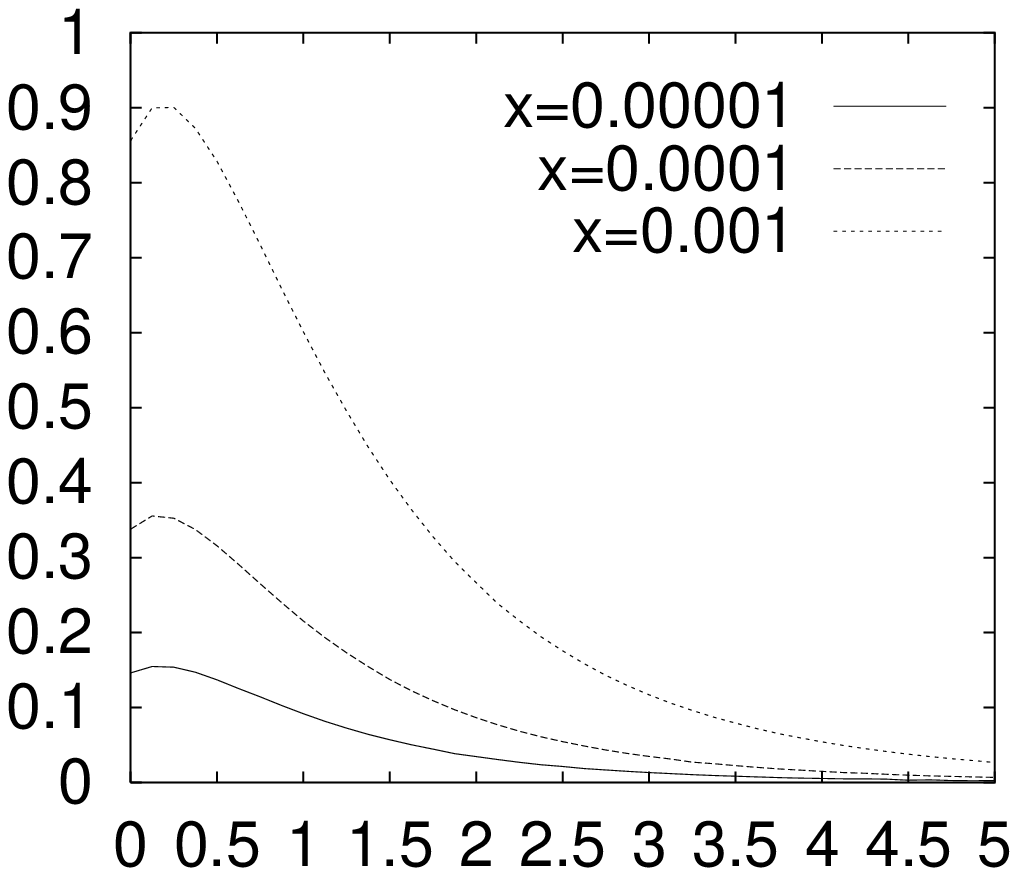}{$R_n$ as a function of $Q^2$ for various
values of $x$ showing that $R_n \rightarrow 0$ as $Q^2 \rightarrow
\infty $. Here $\mu_0^2 = 0.5 ~$GeV$^2$ and $\om_0 = -2.0$
.}{$\log_{10} (Q^2)$}{$R_n$}

\fancyfig{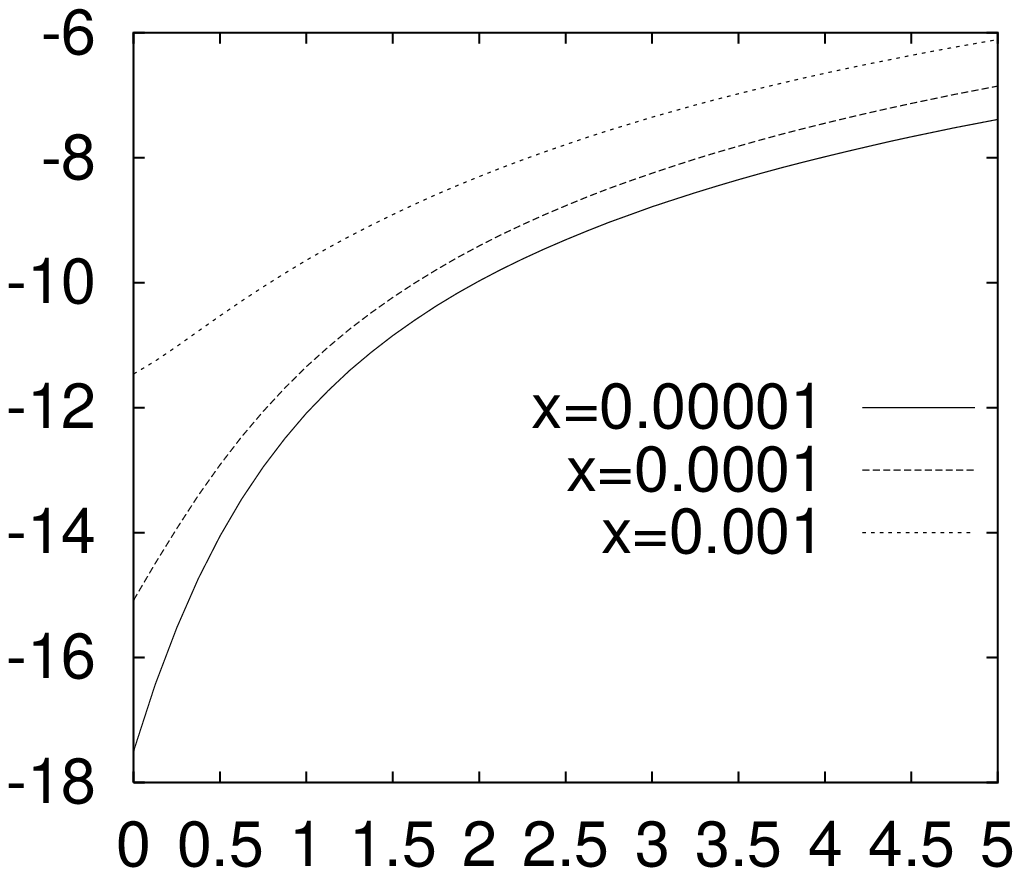}{$R_s$ as a function of $Q^2$ for various
values of $x$. $R_s$ tends to a constant as $Q^2 \rightarrow \infty$.
Here $ \mu_0^2 = 0.5 ~$GeV$^2$, $Q_0^2
= 0.5 $ ~GeV$^2$ and $\om_0 = -2.0$.}{$\log_{10} (Q^2)$}{$R_s$}

\fancyfig{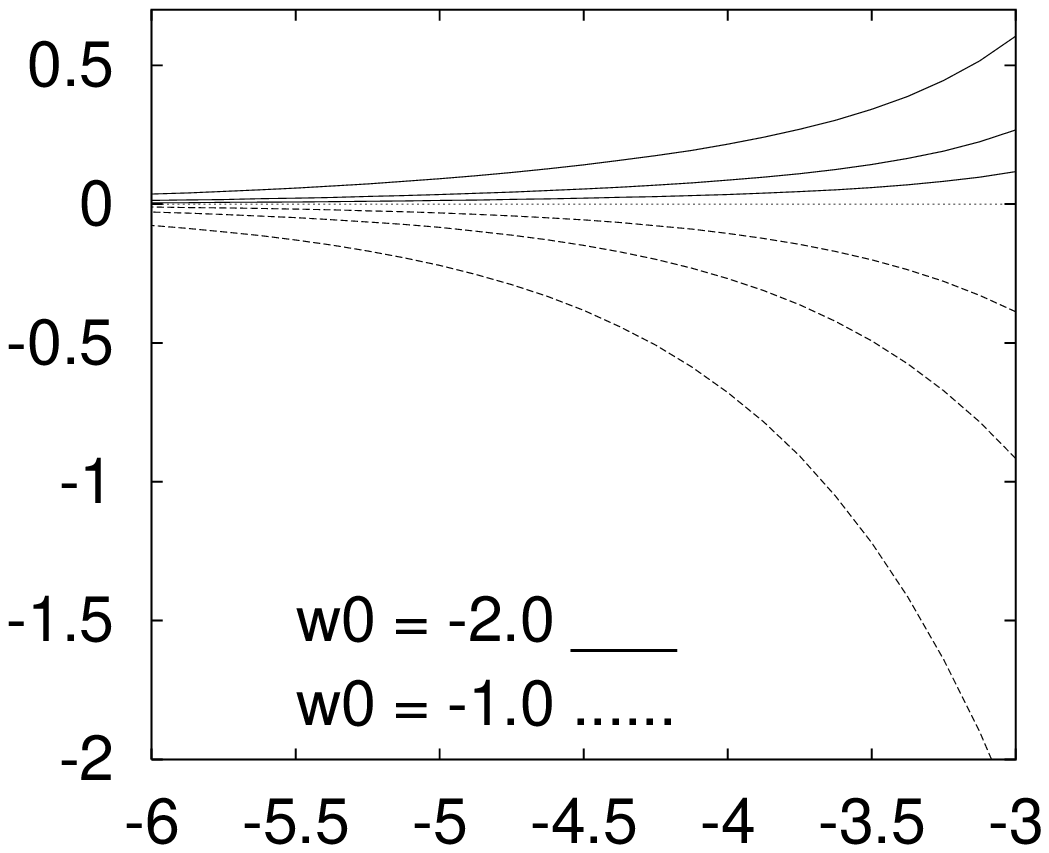}{$R_n$ as a function of $x$ for various
values of $Q^2$. The percentage difference between the two solutions is
less than $ 2 \% $ and tends to zero as $x \rightarrow 0 $.  
The curves shown correspond to (largest first) $Q^2 = 10,100,1000~$
GeV$^2$. Here  $ \mu_0^2 = 0.5 ~$ GeV$^2$.}{$\log_{10} (x)$}{$R_n$}

\fancyfig{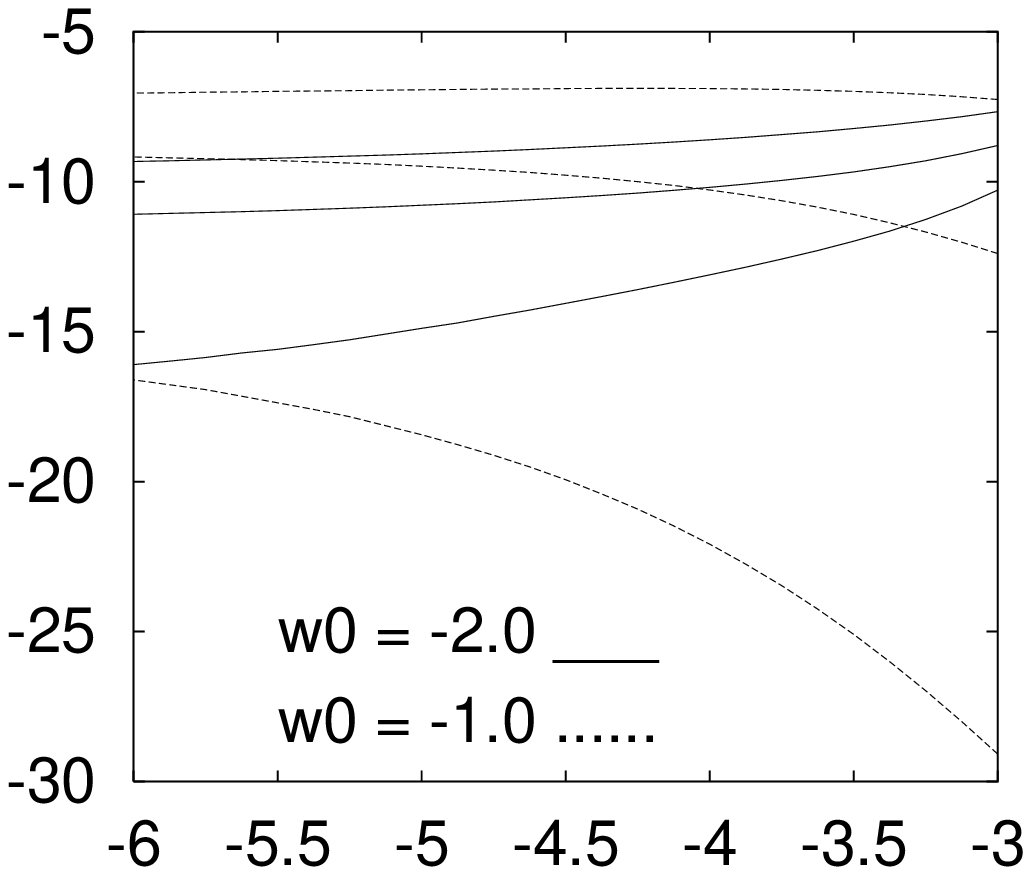}{$R_s$ as a function of  $x$ for various
values of $Q^2$. For the solid curves $\om_0 = -2.0$, wheras for the
dashed curves $\om_0=-1.0$. In each case the curves correspond to
(largest first) $Q^2
= 1,10,100 ~$ GeV$^2$. 
The input parameters are $ \mu_0^2 = 0.5 ~$GeV$^2$,$Q_0^2 =
0.5 ~$ GeV$^2$.}{$\log_{10} (x)$}{$R_s$}


\end{document}